\documentclass[aps,prd,nofootinbib,twocolumn,superscriptaddress,preprintnumbers]{revtex4}

\usepackage{amssymb}
\usepackage{amsmath}
\usepackage{epsfig}

\makeatletter
\def\simgt{\mathrel{\lower2.5pt\vbox{\lineskip=0pt\baselineskip=0pt
           \hbox{$>$}\hbox{$\sim$}}}}
\def\simlt{\mathrel{\lower2.5pt\vbox{\lineskip=0pt\baselineskip=0pt
           \hbox{$<$}\hbox{$\sim$}}}}
\makeatother

\newcommand{\be}{\begin{equation}}
\newcommand{\ee}{\end{equation}}
\newcommand{\bea}{\begin{eqnarray}}
\newcommand{\eea}{\end{eqnarray}}
\newcommand{\Eq}[1]{Eq.~(\ref{#1})}
\newcommand{\Eqs}[2]{Eqs.~(\ref{#1}) and (\ref{#2})}
\newcommand{\Sec}[1]{Sec.~\ref{#1}}
\newcommand{\Secs}[2]{Secs.~\ref{#1} and \ref{#2}}

\newcommand{\vev}[1]{\langle #1 \rangle}

\newcommand{\Lint}{\smallint\!\!{\cal L}}

\begin{document}

\preprint{UCB-PTH-10/16}

\title{Singlet Portal to the Hidden Sector}

\author{Clifford Cheung}
\affiliation{Berkeley Center for Theoretical Physics, 
  University of California, Berkeley, CA 94720, USA}
\affiliation{Theoretical Physics Group, 
  Lawrence Berkeley National Laboratory, Berkeley, CA 94720, USA}

\author{Yasunori Nomura}
\affiliation{Berkeley Center for Theoretical Physics, 
  University of California, Berkeley, CA 94720, USA}
\affiliation{Theoretical Physics Group, 
  Lawrence Berkeley National Laboratory, Berkeley, CA 94720, USA}
\affiliation{Institute for the Physics and Mathematics of the Universe, 
  University of Tokyo, Kashiwa 277-8568, Japan}

\begin{abstract}
Ultraviolet physics typically induces a kinetic mixing between gauge 
singlets which is marginal and hence non-decoupling in the infrared. 
In singlet extensions of the minimal supersymmetric standard model, 
e.g.\ the next-to-minimal supersymmetric standard model, this furnishes 
a well motivated and distinctive portal connecting the visible 
sector to any hidden sector which contains a singlet chiral 
superfield.  In the presence of singlet kinetic mixing, the hidden 
sector automatically acquires a light mass scale in the range 
$0.1~\mbox{--}~100~{\rm GeV}$ induced by electroweak symmetry breaking. 
In theories with $R$-parity conservation, superparticles produced 
at the LHC invariably cascade decay into hidden sector particles. 
Since the hidden sector singlet couples to the visible sector via 
the Higgs sector, these cascades necessarily produce a Higgs boson 
in an order $0.01~\mbox{--}~1$ fraction of events.  Furthermore, 
supersymmetric cascades typically produce highly boosted, low-mass 
hidden sector singlets decaying visibly, albeit with displacement, 
into the heaviest standard model particles which are kinematically 
accessible.  We study experimental constraints on this broad class 
of theories, as well as the role of singlet kinetic mixing in direct 
detection of hidden sector dark matter.  We also present related 
theories in which a hidden sector singlet interacts with the visible 
sector through kinetic mixing with right-handed neutrinos.
\end{abstract}

\maketitle

\section{Introduction}

Physics beyond the standard model has been largely devoted to an 
emerging understanding of the fundamental constituents of matter at 
ever higher energies.  In more recent years, however, some of the focus 
has shifted away from this ``vertical'' line of thinking towards a more 
``horizontal'' perspective concerned with the possibility of hidden 
sectors which are weakly coupled to the visible sector but at the same 
time comprised of particles at observable mass scales.  Indeed, the 
existence of multiple separate sectors is quite plausible in the context 
of string theory, which often predicts a number of geographically 
sequestered sectors~\cite{Giddings:2001yu,Arvanitaki:2009fg,Cheung:2010mc}.

Theories with such light hidden sectors are particularly well motivated 
and exhibit rich phenomenology if there is weak scale supersymmetry. 
With supersymmetry, mass scales of these sectors can be naturally at 
or below the weak scale, since they can be dominantly generated by 
supersymmetry breaking effects induced by interactions with the visible 
sector or ``mandatory'' gravity mediation.  Moreover, supersymmetry can 
offer a unique window into hidden sectors via decay of the lightest 
observable-sector supersymmetric particle (LOSP).  As such, phenomenology 
depends crucially on specific operators connecting visible and hidden 
sector particles.

In general, there may exist heavy mediator particles of mass $M_*$ 
which serve as a bridge between the visible and hidden sectors.  At 
low energies, this typically implies that the two sectors couple only 
through higher dimension operators suppressed by $M_*$.  There are, 
however, two exceptions to this expectation.  First, if the hidden 
sector contains a $U(1)$ gauge field, loops involving heavy mediators 
can generate a marginal operator~\cite{Holdom:1985ag}
\be
  {\cal L} = \chi \int\!d^2\theta\, 
    {\cal W}^\alpha {\cal W}'_\alpha + \textrm{h.c.},
\label{eq:kinmix-gauge}
\ee
where ${\cal W}^\alpha$ and ${\cal W}'_\alpha$ are $U(1)$ 
hypercharge and hidden sector field-strength superfields.  This 
scenario has been extensively studied in literature, for example 
in~\cite{Glashow:1985ud,Pospelov:2007mp,ArkaniHamed:2008qp,Baumgart:2009tn}. 
In this paper we discuss an alternative possibility: if both the visible 
and hidden sectors contain singlet chiral superfields, $S$ and $S'$, 
then a marginal kinetic mixing operator
\be
  {\cal L} = \epsilon \int\!d^4\theta\, S^\dagger S' + \textrm{h.c.},
\label{eq:kinmix-S}
\ee
can persist at low energies, no matter the scale of new physics, $M_*$. 
The size of the coefficient $\epsilon$ is typically a one-loop factor 
or less, $O(\simlt 1/16\pi^2)$.  Note that  {\it any} hidden sector 
which interacts with the visible sector via a marginal operator will 
essentially dictate the phenomenology---the effect of additional sectors 
interacting only through higher dimension operators will be subdominant.

We assume that the visible sector singlet interacts with fields in the 
minimal supersymmetric standard model (MSSM) through a superpotential term
\be
  W = \lambda S H_u H_d,
\label{eq:SHH}
\ee
where $H_u$ and $H_d$ are the up-type and down-type Higgs doublets, 
and $\lambda$ is an $O(\simlt 1)$ coupling.  Indeed, assuming the 
existence of an $R$-parity under which $S$ is even, this is the only 
renormalizable operator which can be written.%
\footnote{The case where $S$ is $R$-parity odd will be discussed in 
 the final section.}
Our analysis will be largely independent of any additional interactions 
involving $S$---a special case is the usual next-to-minimal supersymmetric 
standard model (NMSSM), $\varDelta W = \kappa S^3/3$.

The framework defined by \Eqs{eq:kinmix-S}{eq:SHH} leads to rich and 
distinctive phenomenology.  Three main features are
\begin{list}{\labelitemi}{\leftmargin=1em}
\item
{\bf Spontaneous Scale Generation.}
After electroweak symmetry breaking, singlet kinetic mixing induces an 
effective linear term for $S'$ in the superpotential, set by the scale 
$\Lambda_{\rm eff}^2 = \epsilon (\lambda/2) v^2 \sin 2\beta + \cdots 
\approx O(0.1~\mbox{--}~100~{\rm GeV})$, where $v \equiv \sqrt{\vev{H_u}^2 
+ \vev{H_d}^2}$ and $\tan\beta \equiv \vev{H_u}/\vev{H_d}$.  As a result, 
hidden sector fields generically acquire vacuum expectation values (VEVs) 
of order $\Lambda_{\rm eff}$, yielding light degrees of freedom at this 
scale.  (If the contribution from gravity mediation is larger, the 
characteristic mass scale of the hidden sector may be set by that.)
\item
{\bf Higgs Production with Supersymmetry.}
Since the hidden sector typically contains an $R$-parity odd state which 
is lighter than the LOSP, a superparticle invariably cascades into states 
in the hidden sector.  Because the hidden sector communicates with the 
visible sector only through $S$, which interacts with MSSM states only 
via the Higgs fields, these cascades necessarily produce the Higgs 
boson in an $O(10^{-2}~\mbox{--}~1)$ fraction of events, depending 
on $\Lambda_{\rm eff}$ and the LOSP species.  This leads to a minimum 
rate for high transverse-momentum Higgs production associated with 
significant missing energy.
\item
{\bf Hidden Sector Cascades Return.}
Hidden sector singlets produced by supersymmetric cascades may decay 
back to standard model states, if they are even under $R$-parity. 
Since this process occurs through off-shell Higgs fields, the decay 
product is typically the heaviest possible state which is kinematically 
accessible.  The decay rate scales roughly as $\Gamma \propto 
\epsilon^2 y^2 (m'/v)^2 m'$, where $y$ is the Yukawa coupling of 
the final state and $m'$ the mass of the hidden sector singlet. 
Because of the suppression due to $\epsilon^2$, $y^2$, and $(m'/v)^2$, 
the vertex is generically displaced.  The decay, however, may still 
occur inside the detector for natural values of $\epsilon$, so the 
decay products may be observed at colliders.
\end{list}
These features allow for distinguishing theories with singlet portal 
from alternative scenarios such as $U(1)$ gauge kinetic mixing.

The organization of this paper is as follows.  In \Sec{sec:setup}, we 
describe our basic setup.  We study spontaneous scale generation in 
\Sec{sec:scale-gen}, and interactions between the visible and hidden 
sectors in \Sec{sec:portal}.  In \Secs{sec:portal-in}{sec:portal-out}, 
we describe physics of the ``portal in'' and the ``portal out,'' i.e.\ 
processes converting visible sector states into hidden sector ones and 
vice versa.  We discuss experimental constraints in \Sec{sec:constraints}, 
and study possible implications of this framework on dark matter in 
\Sec{sec:DM}.  Finally, we conclude in \Sec{sec:concl}, and present 
related theories of singlet kinetic mixing in which the hidden sector 
interacts with the visible sector through right-handed neutrinos.

\section{Basic Setup}
\label{sec:setup}

Let us consider a scenario in which there exist two ``separate'' sectors, 
for example those geographically sequestered from each other along 
an extra dimension.  These two sectors may still be connected through 
physics at some high energy $M_*$, e.g.\ at the compactification scale. 
This typically leads to a low energy effective theory in which the two 
sectors interact only through higher dimension operators suppressed 
by $M_*$.

However, if both sectors contain a singlet chiral superfield, $S$ and 
$S'$, then the low energy theory may in general contain the marginal 
kinetic mixing operator in \Eq{eq:kinmix-S}.  For example, this operator 
can be generated by loops of a heavy field $\Phi$/$\bar{\Phi}$ that 
interacts with $S$ and $S'$ through the superpotential
\be
  W = y S \Phi \bar{\Phi} + y' S' \Phi \bar{\Phi}.
\label{eq:kinmix-gen}
\ee
This yields a kinetic mixing operator with the coefficient
\be
  \epsilon \approx \frac{y y'}{16\pi^2} \ln\frac{M_*}{M_\Phi},
\label{eq:epsilon}
\ee
where $M_\Phi$ and $M_*$ are the mass of $\Phi$/$\bar{\Phi}$ and the 
ultraviolet cutoff, respectively.

In general, the precise structure of the heavy-field sector and its 
couplings to $S$ and $S'$ are unknown, so the size of $\epsilon$ is 
model dependent.  It is, however, reasonable to expect that $\epsilon$ 
is of order a one-loop factor or less, and in this paper we mainly 
focus on the range
\be
  10^{-5} \simlt \epsilon \simlt 10^{-1}.
\label{eq:eps-range}
\ee
Since renormalization group evolution from $M_\Phi$ to the weak scale 
does not have a significant effect on the size of $\epsilon$, we consider 
the operator in \Eq{eq:kinmix-S} with \Eq{eq:eps-range} evaluated 
at the weak scale.

\begin{figure}[t]
\begin{center}
  \includegraphics[scale=0.5]{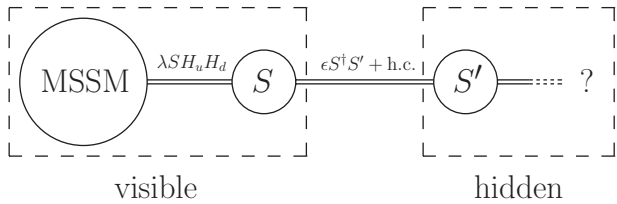}
\end{center}
\caption{A schematic depiction of our setup.  Integrating out heavy 
 states typically induces a marginal kinetic mixing operator between 
 singlet chiral superfields $S$ and $S'$.  The singlet $S$ couples 
 with sizable strength to the MSSM, e.g.\ like in the NMSSM.}
\label{fig:setup}
\end{figure}

As described in the introduction, we assume that $S$ interacts with 
the MSSM states through the interaction in \Eq{eq:SHH}.  We therefore 
consider the following superpotential for the visible sector:
\be
  W = \lambda S H_u H_d + \mu_0 H_u H_d + f(S),
\label{eq:vis-superpot}
\ee
where $f(S)$ is a holomorphic function of $S$.  The conventional NMSSM 
corresponds to $\mu_0 = 0$ and $f(S) = \kappa S^3/3$; but in general 
$\mu_0$ may be of order the weak scale, and $f(S)$ may contain terms 
linear or quadratic in $S$ with weak scale coefficients.  We assume that 
$S$ and $H_{u,d}$ obtain nonvanishing VEVs after supersymmetry breaking, 
which we denote by
\be
  x \equiv \vev{S},
\qquad
  v_u \equiv \vev{H_u},
\qquad
  v_d \equiv \vev{H_d}.
\label{eq:vev-def}
\ee
The supersymmetric mass term for $H_{u,d}$ (the $\mu$ term) is then given 
by $\mu = \mu_0 + \lambda x$.  A schematic depiction of the setup described 
here is given in Fig.~\ref{fig:setup}.

Note that $S$ and $S'$ may not be elementary singlets---the relevant 
renormalizable operators in \Eqs{eq:kinmix-S}{eq:SHH} may exist if there 
are singlets $S$ and $S'$ at the weak scale.  However, if the compositeness 
scales for $S$ and $S'$ are hierarchically smaller than the cutoff scale, 
then the size of $\epsilon$, as well as other interactions of $S$ and 
$S'$, are suppressed accordingly.  Below, we assume that this suppression 
is absent, either because $S$ and $S'$ are elementary or because their 
compositeness scales are sufficiently high.

Finally, let us comment on the possibility that the gauge hierarchy 
might be destabilized in the presence of the visible sector singlet 
$S$.  First, the problem may be avoided if the scale of fundamental 
supersymmetry breaking is sufficiently low, such as in low scale 
gauge mediation.  On the other hand, if the scale of supersymmetry 
breaking is high, dangerous K\"ahler potential operators will 
generically be induced, yielding an unacceptably large tadpole for 
$S$~\cite{Polchinski:1982an}.  This problem can be solved if the 
theory possesses a (softly broken) discrete symmetry, or if anthropic 
selection plays a role in determining the weak scale.  Because 
hierarchy destabilization is a generic pitfall of singlet extensions 
of the MSSM, we will not address it further in this paper.

\section{Spontaneous Scale Generation}
\label{sec:scale-gen}

The operator in \Eq{eq:kinmix-S} spontaneously generates a scale of 
$O(0.1~\mbox{--}~100~{\rm GeV})$ in the hidden sector.  To see this, 
let us denote the component fields of $S$ and $S'$ by
\bea
  S  &=& s + \sqrt{2} \theta \tilde s + \theta^2 F_{S},
\label{eq:S-component}\\
  S' &=& s' + \sqrt{2} \theta \tilde s' + \theta^2 F_{S'},
\label{eq:S'-component}
\eea
and expand \Eq{eq:kinmix-S} as
\be
  {\cal L}_{\rm kin} = \epsilon( -\partial^\mu s^\dagger \partial_\mu s' 
  + i \tilde{s}^\dagger \bar{\sigma}^\mu \partial_\mu \tilde{s}' 
  + F_S^\dagger F_{S'}) + {\rm h.c.}
\label{eq:kinmix-expand}
\ee
After electroweak symmetry breaking, $F_S^\dagger$ acquires a VEV 
which induces a tadpole for $F_{S'}$.  This term is equivalent to 
adding an effective Polonyi term to the hidden sector superpotential, 
\be
  W_{\rm eff} =  -\Lambda_{\rm eff}^2 S',
\label{eq:eff-Pol}
\ee
where
\bea
  \Lambda_{\rm eff}^2 &\equiv& -\epsilon \vev{F_S^\dagger} 
  = \epsilon \left( \frac{\lambda v^2 \sin 2\beta}{2} 
    + \frac{df(x)}{dx} \right)
\nonumber\\
  &\approx& O(0.1~\mbox{--}~100~{\rm GeV})^2,
\label{eq:effpol}
\eea
corresponding to the range of $\epsilon$ quoted in \Eq{eq:eps-range} 
and $1 \simlt \tan\beta \simlt 50$.  Here the function $f$ was defined 
in \Eq{eq:vis-superpot} and we have made the reasonable assumption 
that $df/dx$ is not much greater than the weak scale.

Note that if the $df/dx$ term is a subdominant contribution to 
$\Lambda_{\rm eff}^2$, then $\Lambda_{\rm eff}^2 \propto \sin 2\beta$ 
and may be significantly suppressed at large $\tan\beta$.  This 
is an important difference from the case of gauge kinetic mixing 
of hypercharge and a hidden sector $U(1)'$.  There electroweak 
symmetry breaking induces an effective Fayet-Iliopoulos term for 
the $U(1)'$ gauge field which goes as $\xi \propto \cos 2\beta$ 
and is thus largely insensitive to $\tan\beta$ unless $\tan\beta 
\approx 1$~\cite{Baumgart:2009tn}.

The effective Polonyi term in \Eq{eq:effpol} injects the scale 
$\Lambda_{\rm eff}$ into the hidden sector.  As a result, the masses 
of hidden sector fields are typically of this order.  For instance, 
consider a simple hidden sector theory in which $S'$ has a trilinear 
superpotential interaction.  Together with \Eq{eq:eff-Pol}, the 
effective hidden sector superpotential is then
\be
  W_{\rm hid} = -\Lambda_{\rm eff}^2 S'+ \frac{\kappa'}{3} S'^3.
\label{eq:simple-W}
\ee
The scalar potential is
\be
  V_{\rm hid} = |\kappa' s'^2 - \Lambda_{\rm eff}^2|^2.
\label{eq:simple-V}
\ee
Thus, $x' \equiv \vev{S'} = \sqrt{\Lambda_{\rm eff}^2/\kappa'}$, and 
the vacuum aligns to preserve supersymmetry.%
\footnote{Despite the presence of an effective Polonyi term, the 
 vacuum typically adjusts to preserve supersymmetry.  A notable exception 
 is O'Raifeartaigh-like constructions, such as the one defined by 
 $W_{\rm hid} = -\Lambda_{\rm eff}^2 S' + \mu' T' U' + \lambda' S' T'^2$.}
Furthermore, $s'$ and $\tilde s'$ both acquire a mass $m'^2 = 
4|\kappa' \Lambda_{\rm eff}^2|$.  For $O(\simlt 1)$ values of $\kappa'$, 
this implies a hidden sector singlet mass in the range
\be
  m' \approx O(10~{\rm MeV}~\mbox{--}~100~{\rm GeV}).
\label{eq:mass-range}
\ee
The spontaneous scale generation exhibited by this simple model is 
a generic feature of models with kinetically mixed singlets.

Of course, $m'$ can exceed $\Lambda_{\rm eff}$ if the hidden sector 
has additional sources of mass generation.  In particular, this may 
occur if $W_{\rm hid}$ contains explicit mass terms or if the hidden 
sector receives large supersymmetry breaking contributions, for 
example from gravity mediation.  These effects are highly model 
dependent---explicit mass terms are easily forbidden by any number 
of chiral, $R$, or discrete symmetries, and the scale of supersymmetry 
breaking may be low, in which case gravity mediated contributions 
will be subdominant.  Nevertheless, as to be as model independent 
as possible, the remainder of our discussion will be agnostic 
about the origin of $m'$, and will consider the possibility 
that $m'$ may be as large as the weak scale, regardless of 
the value of $\Lambda_{\rm eff}$.

\section{The Portal}
\label{sec:portal}

We now discuss the effective interactions between visible and hidden 
sector fields.  To simplify the discussion, we consider only a single 
field $S'$ in the hidden sector---the existence of possible additional 
fields will not affect our basic conclusions.  The most general 
hidden sector superpotential is then written as
\be
  W_{\rm hid} = \frac{m'}{2} S'^2 + \frac{\kappa'}{3} S'^3,
\label{eq:W_hid}
\ee
because we can always define the origin of $S'$ so that the linear term 
in the superpotential vanishes (unless $\partial^2 W_{\rm hid}/\partial 
S'^2 = 0$ in the original basis).  Note that $W_{\rm hid}$ includes 
the effect of the Polonyi term in \Eq{eq:eff-Pol}, as in \Eq{eq:simple-W}. 
For the model of \Eq{eq:simple-W}, for example, \Eq{eq:W_hid} 
is obtained after the shift $S' \rightarrow S' + x'$, so that 
$m' = 2 \sqrt{\kappa'} \Lambda_{\rm eff}$.

In what follows, we will assume $m', \kappa' \neq 0$, which indeed 
represents the situation for generic hidden sectors.  We will mostly 
ignore supersymmetry breaking effects in the hidden sector, which is 
typically a good approximation.  (It is indeed a very good approximation 
if the dominant superparticle masses arise from gauge mediation in the 
visible sector.)  The case where the supersymmetry breaking effects 
are important will be discussed briefly.

In general, interactions between the visible and hidden sector fields can 
be obtained by canonically normalizing fields, starting from the original 
Lagrangian containing kinetic mixing terms of \Eq{eq:kinmix-expand}. 
For small $\epsilon$, however, there is a simple way to obtain the leading 
interactions between the two sectors, which we will present below.

Let us first consider the hidden sector fermion, $\tilde{s}'$.  At the 
leading order in $\epsilon$, the kinetic mixing between $\tilde{s}$ and 
$\tilde{s}'$ in ${\cal L}_{\rm kin}$ can be removed by the shift
\be
  \tilde{s}' \rightarrow \tilde{s}' - \epsilon \tilde{s}.
\label{eq:singlino-shift}
\ee
This induces interactions between visible and hidden sector fields, 
$-\epsilon \tilde{s} (\partial /\partial \tilde{s}') {\cal L}_{\rm hid}$, 
where ${\cal L}_{\rm hid}$ denotes the hidden sector interaction 
Lagrangian.  For the theory defined in \Eq{eq:W_hid}, the resulting 
term is $2 \epsilon \kappa' s' \tilde{s}' \tilde{s}$.  Note that 
an interaction term generated in this way always contains only 
a single visible sector field.

Another important effect of the shift in \Eq{eq:singlino-shift} is to 
induce a mass mixing, $\epsilon m' \tilde{s}' \tilde{s}$.  As a result, 
the fermion mass matrix takes the following schematic form
\be
  {\cal M}_{\rm fermion} = 
    \left( \begin{array}{ccc|c}
      & & & \\
      & m \approx \textrm{weak scale} & & \\
      & & &  \epsilon m' \\ \hline
      \;\;\;\; & & \epsilon m' & m'
    \end{array} \right),
\label{eq:M_fermion}
\ee
where the upper-left block corresponds to the neutralino mass matrix 
of the visible sector in the $\{\tilde{b}, \tilde{w}, \tilde{h}_u, 
\tilde{h}_d, \tilde{s} \}$ basis, and the bottom-right block corresponds 
to $\tilde{s}'$.  Here $m$ broadly denotes quantities which are roughly 
of order the weak scale.  After diagonalizing ${\cal M}_{\rm fermion}$, 
it is clear that mixing angles of $\tilde{s}'$ into visible sector 
fermions go as
\be
  \theta_{\tilde{s}' \tilde{b}} \sim \theta_{\tilde{s}' \tilde{w}} 
    \sim \theta_{\tilde{s}' \tilde{h}_u} 
    \sim \theta_{\tilde{s}' \tilde{h}_d} 
    \sim \theta_{\tilde{s}' \tilde{s}} \sim \epsilon \frac{m'}{m},
\label{eq:singlino-mix}
\ee
up to $O(\simlt 1)$ coefficients which are model dependent.%
\footnote{To be precise, the $\theta$ parameters here represent the 
 fractions of $\tilde{b}, \tilde{w}, \tilde{h}_u, \tilde{h}_d, \tilde{s}$ 
 which contain the ``mostly $\tilde{s}'$ mass eigenstate,'' which is 
 purely $\tilde{s}'$ at the leading order in $\epsilon$.}
These mixings induce interaction terms which involve more than one 
visible sector field.  For example, if the visible sector superpotential 
contains a term $\kappa S^3/3$, then this mixing leads to an interaction 
$-2 \theta_{\tilde{s}' \tilde{s}} \kappa s \tilde{s} \tilde{s}'$.

We next consider the hidden sector scalars.  As in the case of fermions, 
we can remove the kinetic mixing between $s$ and $s'$ at the leading order 
in $\epsilon$ via the shift
\be
  s' \rightarrow s' - \epsilon s.
\label{eq:singlet-shift}
\ee
This induces interaction terms $-\epsilon s (\partial/\partial s') 
{\cal L}_{\rm hid}$, each of which contains only a single visible 
sector field.

Interactions involving more than one visible sector scalar predominantly 
arise from the $F_S^\dagger F_{S'}$ term in \Eq{eq:kinmix-expand}. 
By expanding both $F_S^\dagger$ and $F_{S'}$ to first order in field 
fluctuations, we find
\be
  \mathcal{L}_{\rm kin} \supset 
    \epsilon \left(\lambda v(h_u \cos \beta + h_d \sin\beta) 
    + \frac{d^2 f(x)}{dx^2} s\right)(m's')^\dagger,
\label{eq:Lmix-scalar}
\ee
which mixes $h_u$, $h_d$, and $s$ with $s'$ with coefficients of order 
$\epsilon m m'$.  Consequently, the scalar mass-squared matrix is 
schematically
\be
  {\cal M}^2_{\rm scalar} \approx 
    \left( \begin{array}{ccc|c}
      & & & \\
      & m^2 \approx (\textrm{weak scale})^2 & & \epsilon m m' \\
      & & & \\ \hline
      \;\;\; & \epsilon m m' & \;\;\; & m'^2
    \end{array} \right),
\label{eq:M_scalar}
\ee
where the upper-left block corresponds to the neutral Higgs fields of 
the visible sector, and the bottom-right block corresponds to $s'$. 
Here, the basis of scalars is spanned by both CP even and odd components. 
In the case that CP is conserved, ${\cal M}^2_{\rm scalar}$ decomposes 
into CP even and CP odd blocks.  Interestingly, we find that the mixing 
angles of $s'$ with the visible sector states scale as in the fermion 
sector:
\be
  \theta_{s' h_u} \sim \theta_{s' h_d} \sim \theta_{s' s} 
    \sim \epsilon \frac{m'}{m},
\label{eq:singlet-mix}
\ee
where $h_u$, etc., collectively denote the CP even and CP odd components. 
Note that $s'$ does not mix into (the longitudinal component of) 
the $Z$ boson as dictated by gauge invariance, which implies that 
$\theta_{s' h_u}/\theta_{s' h_d} \propto \cot\beta$ for the CP odd 
component of $s'$.

We finally comment on possible effects of supersymmetry breaking.  If 
the scale of supersymmetry breaking in the hidden sector, $\tilde{m}'$, 
is larger than the scale $m'$ in \Eq{eq:W_hid}, then the mass scale of 
the hidden sector will be set by $\tilde{m}'$ (at least for the scalars). 
Moreover, mixing angles between $s'$ and visible sector scalars receive 
an extra contribution of order
\be
  \delta\theta_{s' h_u} \sim \delta\theta_{s' h_d} \sim \delta\theta_{s' s} 
    \sim \epsilon \left( \frac{\tilde{m}'}{m} \right)^2,
\label{eq:singlet-mix-2}
\ee
since the soft supersymmetry breaking mass-squared matrix obtains 
a nonvanishing $s$-$s'$ component of order $\epsilon \tilde{m}'^2$ 
after the shift of \Eq{eq:singlet-shift}.  For $\tilde{m}' \gg m'$, 
this contribution may be larger than that in \Eq{eq:singlet-mix}.

To summarize, we find that the portal between the visible and hidden 
sectors takes the form
\be
  {\cal L}_{\rm portal} = {\cal L}_\textrm{portal in} 
    + {\cal L}_\textrm{portal out} + {\rm h.c.},
\label{eq:L_portal}
\ee
where
\begin{align}
  {\cal L}_\textrm{portal in} 
  &= \left( -\epsilon \tilde{s} \frac{\partial}{\partial \tilde{s}'} 
    - \epsilon s \frac{\partial}{\partial s'} \right) {\cal L}_{\rm hid},
\label{eq:portal-in}\\
  {\cal L}_\textrm{portal out} 
  &= \left( \tilde{s}' \sum_{\tilde{\phi}} 
    \theta_{\tilde{s}'\tilde{\phi}} \frac{\partial}{\partial \tilde{\phi}} 
    + s' \sum_\phi \theta_{s'\phi} \frac{\partial}{\partial \phi} \right) 
    {\cal L}_{\rm vis}.
\label{eq:portal-out}
\end{align}
Here, ${\cal L}_{\rm hid}$ and $ {\cal L}_{\rm vis}$ denote interaction 
Lagrangians of the hidden and visible sectors, respectively, while 
$\tilde{\phi}$ and $\phi$ run over the visible sector neutralinos 
and neutral Higgs states, respectively.%
\footnote{The second term of \Eq{eq:portal-out} is only schematic, as 
 the mixing angles for CP even and odd components differ in general. 
 For the CP-conserving case, the precise expression is given by
 \be
    {\cal L}_\textrm{portal out} \supset 
    \left( s'_R \sum_{\phi_R} \theta_{s'_R \phi_R} 
      \frac{\partial}{\partial \phi_R} 
    + s'_I \sum_{\phi_I} \theta_{s'_I \phi_I} 
      \frac{\partial}{\partial \phi_I} \right) {\cal L}_{\rm vis},
 \nonumber
 \ee
 where $s' = (s'_R + i s'_I)/\sqrt{2}$, and $\phi_R$ and $\phi_I$ represent 
 real and imaginary components of the visible sector Higgs fields, 
 respectively.  In the general case with CP violation, both of the sums 
 run over $\phi_R$ and $\phi_I$, i.e.\ both $s'_R$ and $s'_I$ mix with 
 all the visible sector Higgs fields.}
The mixing angles $\theta_{\tilde{s}'\tilde{\phi}}$ and $\theta_{s'\phi}$ 
are given by \Eqs{eq:singlino-mix}{eq:singlet-mix}, and are all of 
order $\epsilon m'/m$ (unless the contribution of \Eq{eq:singlet-mix-2} 
is larger).  As we will see, supersymmetric cascades at colliders will 
mainly portal in to the hidden sector via ${\cal L}_\textrm{portal in}$ 
and portal out of the hidden sector via ${\cal L}_\textrm{portal out}$. 
Thus, these processes are controlled by interaction terms with 
coefficients of order $\epsilon$ and $\epsilon m'/m$, respectively.

\section{To the Hidden Sector}
\label{sec:portal-in}

In this section, we discuss the collider signatures associated with 
the decay of the LOSP into the hidden sector via the singlet portal. 
In theories with $R$-parity conservation, superparticles produced 
at colliders will cascade down to lighter $R$-parity odd particles. 
Since the hidden sector typically contains an $R$-parity odd state 
lighter than the LOSP, these cascades produce hidden sector particles.

We first consider the case in which the LOSP is the lightest neutralino. 
Since the bino or wino does not couple directly to the hidden sector, 
the singlino and Higgsino components are most relevant.

The singlino component leads to an invisible decay $\tilde{s} 
\rightarrow x' \tilde{x}'$ through an $\epsilon$-suppressed coupling 
in \Eq{eq:portal-in}, where $x'$ and $\tilde{x}'$ are hidden sector 
particles to which $\tilde{s}'$ couples with sizable strength.  It 
also leads to a subdominant decay mode $\tilde{s} \rightarrow h \tilde{x}', 
s \tilde{x}'$ through an $\epsilon m'/m$-suppressed coupling in 
\Eq{eq:portal-out}, where $h$ represents either a neutral Higgs or 
$Z$ boson.  In particular, this leads to the (standard model like) 
Higgs boson in the final state of the $\tilde{s}$ decay with a probability 
of $O(m'^2/m^2)$.  Note that the singlino is produced only through 
the interaction of \Eq{eq:SHH}, so that the Higgs boson is also 
produced with a probability of $O(\simgt 1/16\pi^2)$ when a heavier 
superparticle decays into the singlino.

\begin{figure}[t]
\begin{center}
  \includegraphics[scale=0.55]{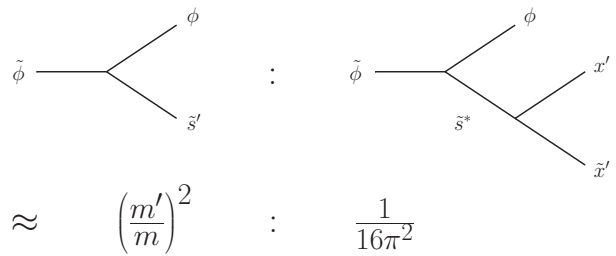}
\end{center}
\caption{Decays of the Higgsino, squark, or slepton LOSP, represented 
 collectively by $\tilde{\phi}$.  Here $\phi$ denotes a Higgs (or 
 electroweak gauge) boson, quark, or lepton, while $x'$ and $\tilde{x}'$ 
 denote hidden sector fields to which $\tilde{s}'$ couples with 
 sizable strength.}
\label{fig:decays}
\end{figure}

The Higgsino component, on the other hand, leads to either a two-body 
decay through $\tilde{h} \rightarrow h \tilde{s}', s \tilde{s}'$ via 
an $\epsilon m'/m$-suppressed coupling, or a three-body decay through 
$\tilde{h} \rightarrow h \tilde{s}^* \rightarrow h x' \tilde{x}'$ via 
an off-shell $\tilde{s}$ and an $\epsilon$-suppressed coupling.  These 
processes have competitive rates, with a ratio $\Gamma_{\tilde{h} 
\rightarrow h \tilde{s}', s \tilde{s}'}/\Gamma_{\tilde{h} \rightarrow 
h x' \tilde{x}'} \approx 16\pi^2 m'^2/m^2$, which depends strongly on 
the size of $m'$; see Fig.~\ref{fig:decays}.  Note that decay through 
the Higgsino component always leads to the Higgs boson with an $O(1)$ 
probability, regardless of the size of $m'$.

We next consider the case of a chargino LOSP.  As before, the charged 
wino component does not couple directly to the hidden sector, so that 
only the charged Higgsino component is relevant.  Similarly to the neutral 
component, the charged Higgsino decays as $\tilde{h}^\pm \rightarrow 
h^\pm \tilde{s}', h^\pm x' \tilde{x}'$ with competitive rates, where 
$h^\pm$ represents either a charged Higgs or $W$ boson.

We now consider the case in which the LOSP is a squark, slepton, or 
sneutrino.  In this case, the LOSP decays to a quark, lepton or neutrino, 
plus invisible decay products, as depicted in Fig.~\ref{fig:decays}. 
If the Yukawa couplings of the LOSP are large, then the LOSP decays 
to an off-shell Higgsino which then mixes into a hidden sector singlino 
$\tilde{s}'$, or an off-shell singlino $\tilde{s}^*$ decaying into 
hidden sector states $x'\tilde{x}'$.  Alternatively, if the LOSP Yukawa 
couplings are small, then the LOSP decays to an off-shell gaugino 
which is converted to the Higgsino and then to either $\tilde{s}'$ 
or $x'\tilde{x}'$.  Either way, the rates to $\tilde{s}'$ and 
$x'\tilde{x}'$ are competitive, with the ratio again given by 
$\approx 16\pi^2 m'^2/m^2$.

If the LOSP is the gluino or (almost) pure bino or wino, then it 
decays through an off-shell sfermion.  The final state is then the 
same as the corresponding sfermion decay, with an extra quark, lepton, 
or neutrino.

In summary, the above analysis highlights a number of salient points. 
First, the visible products of a supersymmetric cascade can be different 
from conventional supersymmetric theories.  For example, if the LOSP is 
the lightest neutralino in which the Higgsino fraction is larger than 
the singlino one, then its decay leads to the Higgs boson with an $O(1)$ 
fraction of the time, even if the LOSP is not Higgsino-like.  This leads 
to a distinct signature in which an $O(1)$ fraction of the supersymmetric 
events is accompanied by the Higgs boson.  While it is possible to 
mimic this in a conventional scenario, e.g.\ by having the Higgsino-like 
next-to-lightest supersymmetric particle decaying into the gravitino, 
observing many Higgs bosons may be an important first step in identifying 
the present scenario.

Second, since MSSM states interact with the hidden sector only through 
the Higgs sector, supersymmetric cascades are required to pick up 
a Higgs VEV or emit a physical (neutral or charged) Higgs boson (or 
the corresponding longitudinal electroweak gauge boson).  Given that 
the existence of cascades containing a charged Higgs boson typically 
implies the existence of cascades containing a neutral Higgs boson,%
\footnote{An exception to this arises in a special case in which a light 
 top squark can decay only to the bottom and a charged Higgsino, which 
 in turn decays into an (off-shell) charged Higgs boson.  There is 
 then no corresponding process which yields a neutral Higgs boson.}
we should expect a Higgs boson in the visible products of supersymmetric 
cascades with
\be
  \frac{\mbox{\# of SUSY events with }h}{\mbox{\# of SUSY events}}
  \approx O(10^{-2}~\mbox{--}~1),
\label{eq:Ph}
\ee
where the Higgs boson is typically produced at the end of the visible 
sector cascade.  This is because any cascade decay process involving 
a Higgs VEV is necessarily accompanied by the corresponding subleading 
process in which the Higgs VEV is replaced by a physical on-shell Higgs 
boson, which is suppressed by an additional $1/16\pi^2$ phase space factor. 
This implies that there is a minimum rate for high transverse-momentum 
Higgs production associated with significant missing energy, which 
may help to discover the Higgs boson through the $b\bar{b}$ decay 
mode~\cite{Kribs:2009yh}.

Finally, note that LOSP decays will produce hidden sector scalars ($s'$ 
or $x'$) in a significant fraction of events, $O(\simgt 1/16\pi^2)$. 
Indeed, even if the LOSP has a dominant branching fraction to $\tilde{s}'$, 
there is typically a competitive decay mode to $x'\tilde{x}'$ through 
an off-shell $\tilde{s}$.  This fact can have a significant implications 
for the ``portal out'' of the hidden sector discussed in the next section.

\section{From the Hidden Sector}
\label{sec:portal-out}

As we have seen, the characteristic mass scale of the hidden sector is 
less than or of order the weak scale.  In fact, the dynamically induced 
scale $\Lambda_{\rm eff}$ in \Eq{eq:effpol} may be significantly smaller 
than the weak scale due to $\sin 2\beta$ suppression.  Here we assume 
that the hidden sector scale is indeed below the superpartner threshold. 
In this case, hidden sector states will be produced in supersymmetric 
cascades, and may return via decays into standard model particles.

Since the hidden sector couples to the visible sector through the Higgs 
sector, the dominant final state is generically the heaviest standard 
model particles which are kinematically accessible.  Whether these 
return processes indeed occur at colliders may depend on the spectrum 
of the hidden sector.  For example, if cascades produce only hidden 
sector states which are stable, e.g.\ the lightest supersymmetric 
particle, then there will be no return process.  However, as we have 
seen in the previous section, supersymmetric cascades typically 
produce hidden sector scalars with a significant fraction, which in 
turn decay back to the standard model.  In particular, in the minimal 
theory defined in \Sec{sec:portal}, $s'$ scalars are directly produced 
via portal in.  Since $s'$ couples to the standard model through mixing 
with the Higgs field, it necessarily decays to standard model particles.%
\footnote{If $s'$ is heavier than $\tilde{s}'$, then $s'$ may decay 
 into $\tilde{s}'$ and the gravitino.  For gravity mediation, the decay 
 rate is very small $\Gamma_{s' \rightarrow \tilde{s}' \tilde{G}} 
 \simeq m^5/16\pi F^2$, where $\sqrt{F}$ is the scale of 
 fundamental supersymmetry breaking.  For gauge mediation, 
 $s'$ and $\tilde{s}'$ are nearly degenerate, $\delta m' \approx 
 {\rm max}\{ \epsilon^2 m', \epsilon^2 m^2/16\pi^2 m' \} \ll m'$, 
 so that the decay rate $\Gamma_{s' \rightarrow \tilde{s}' \tilde{G}} 
 \simeq m' \delta m'^4/\pi F^2$ is again suppressed.  In either 
 case, the partial decay rate to the gravitino is much smaller 
 than the dominant one to standard model particles.}

Let us now consider decay of the $s'$ scalars.  For definiteness, we 
assume that the Higgs sector preserves CP and consider the CP even 
component (real part) of $s'$.  We assume that the mass of $s'$, 
which we denote here by $m'$ regardless of its origin, is below 
$100~{\rm GeV}$; the case $m' \simgt 100~{\rm GeV}$ will be discussed 
later.  The terms relevant to the decay are then
\be
  {\cal L}_\textrm{portal out} 
  \supset s' \left( \theta_{s' h_u} \frac{\partial}{\partial v_u} 
    + \theta_{s' h_d} \frac{\partial}{\partial v_d} \right) 
    {\cal L}_{\rm SM}.
\ee
Here we have defined
\bea
  {\cal L}_{\rm SM} &=& -\frac{1}{4 e^2(v_u,v_d)} F_{\mu\nu} F^{\mu\nu} 
    - \frac{1}{4 g_{\rm s}^2(v_u,v_d)} G_{\mu\nu}^a G^{a\mu\nu}
\nonumber\\
  && {} - v_u \left( \sum_i y_{u_i} \bar{u}_i u_i \right)
\nonumber\\
  && {} - v_d \left( \sum_i y_{d_i} \bar{d}_i d_i 
    + \sum_i y_{\ell_i} \bar{\ell}_i \ell_i \right),
\eea
where $e$ and $g_{\rm s}$ are the electromagnetic and QCD couplings 
renormalized at $m'$, and the sum over up-type quarks $u_i$, down-type 
quarks $d_i$, and charged leptons $\ell_i$ runs over states which are 
kinematically allowed in the $s'$ decay.  As expected, $s'$ couples to 
standard model fermions through Yukawa couplings, and to gauge bosons 
through one-loop renormalization effects from heavy states, whose 
masses depend on $v_u$ and $v_d$.

When $m'$ is above the QCD scale $\Lambda_{\rm QCD}$, it is reasonable 
to compute decay rates to gluons and quarks at the partonic level 
using ${\cal L}_{\rm SM}$.  However, for $m' \simlt \Lambda_{\rm QCD}$, 
$s'$ no longer decays to constituent partons but to hadrons.  To 
estimate the hadronic branching ratio in this mass range, we replace 
the terms involving $g/u/d/s$ in ${\cal L}_{\rm SM}$ with the $SU(3)_f$ 
chiral Lagrangian describing the dynamics of octet mesons.  We 
interpolate between the partonic theory and the chiral Lagrangian 
at the charm threshold $m' = 2m_c$.  See~\cite{Chivukula:1989ze} 
for the details of this calculation.

\begin{figure}[t]
\begin{center}
  \includegraphics[scale=1.0]{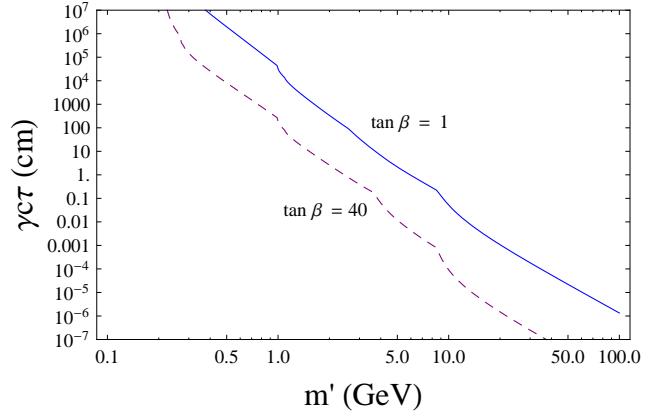}
\end{center}
\caption{Decay length $\gamma c\tau$ of the hidden sector singlet $s'$ 
 as a function of its mass $m'$.  Here the solid (blue) and dashed 
 (purple) lines correspond to $\tan\beta = 1$ and $40$, respectively. 
 For concreteness, we have used a boost factor $\gamma = m /m'$ and 
 mixing angles $\theta_{s' h_u} = \theta_{s' h_d} = \epsilon m'/m$ 
 where $m = 300~{\rm GeV}$ and $\epsilon = 10^{-2}$.}
\label{plot:displacement}
\end{figure}
Having obtained the couplings of $s'$ to standard model fields, 
we can now compute the decay length and branching ratios of $s'$. 
In Fig.~\ref{plot:displacement}, we show the decay length $c\tau$ 
multiplied by a boost factor $\gamma$ as a function of $m'$. 
For illustrative purposes, we have taken $\gamma = m /m'$ 
and $\theta_{s' h_u} = \theta_{s' h_d} = \epsilon m'/m$ where 
$m = 300~{\rm GeV}$ and $\epsilon = 10^{-2}$.  (Of course, quantities 
represented by $m$ in $\gamma$, $\theta_{s' h_u}$ and $\theta_{s' h_d}$ 
are not the same; $\gamma$ even varies event by event.)  The scaling 
of the decay length with respect to these parameters is given by 
$\gamma c\tau \propto \gamma \theta_{s' h_{u,d}}^{-2} \propto 
\epsilon^{-2} (m/m')^3$.

\begin{figure}[t]
\begin{center}
  \includegraphics[scale=1.0]{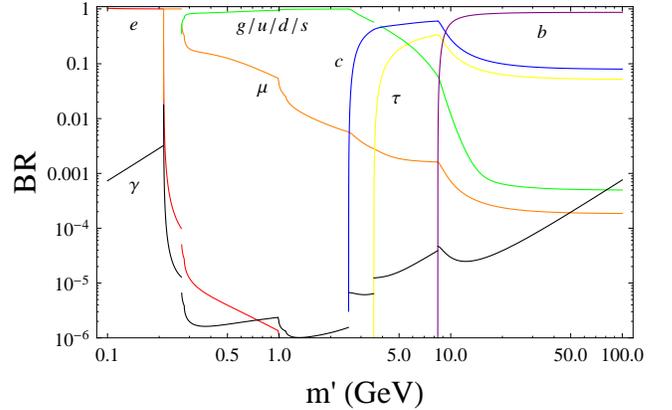}
\end{center}
\caption{Branching ratios of the hidden sector singlet $s'$ into 
 $\{e,\mu,\tau,g/u/d/s,c,b,\gamma\}$, corresponding to the \{red, orange, 
 yellow, green, blue, purple, black\} lines, as a function of $m'$.  Here, 
 we have taken $\theta_{s' h_u} = \theta_{s' h_d}$ and $\tan\beta = 1$. 
 Below the $2m_c$ threshold, decays to partonic $g/u/d/s$ are replaced 
 by decays to octet mesons.}
\label{plot:BR1}
\end{figure}
\begin{figure}[t]
\begin{center}
  \includegraphics[scale=1.0]{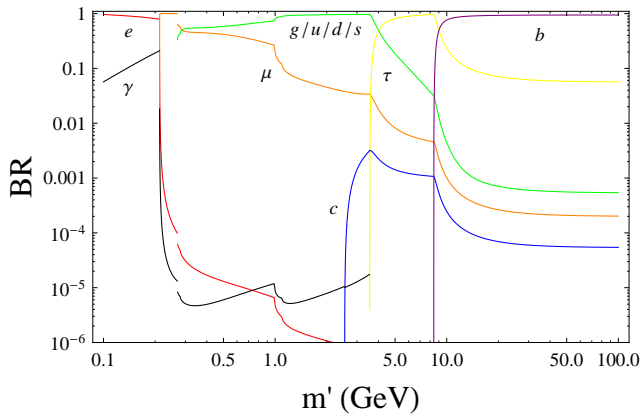}
\end{center}
\caption{The same plot as in Fig.~\ref{plot:BR1} except for 
 $\tan\beta = 40$.}
\label{plot:BR40}
\end{figure}
The branching ratios for $s'$ decay as a function of $m'$ are shown in 
Figs.~\ref{plot:BR1} and \ref{plot:BR40} for $\tan\beta = 1$ and $40$, 
respectively.  Here, we have taken $\theta_{s' h_u}/\theta_{s' h_d} = 1$; 
all the dependencies on other free parameters cancel in the branching 
ratios.  We can see that $s'$ decays generically to the heaviest possible 
state kinematically available, although there are some exceptions, 
e.g.\ see $2m_\tau < m' < 2m_b$ for small $\tan\beta$.  The dependence 
on $\tan\beta$ appears clearly in the leptonic branching ratios for 
$m' > 2 m_\mu$, which arises from the fact that the ratios of the Yukawa 
couplings $y_{u_i}/y_{\ell_i}$ depend on $\tan\beta$.  Note that these 
branching ratios, however, are not uniquely fixed by $\tan\beta$; they 
also depend on $\theta_{s' h_u}/\theta_{s' h_d}$ which we have taken 
to be unity for illustrative purposes.  The rare branching ratio 
into photons is also strongly affected by $\tan\beta$.

Figure~\ref{plot:displacement} shows that for $m' \simlt O(10~{\rm GeV})$, 
$s'$ is long-lived in collider timescales, leading to a displaced decay 
vertex from which standard model particles originate.  The decay product 
typically consists of two particles, e.g.\ $e^+e^-$, $\mu^+\mu^-$, or 
$\pi^+\pi^-$, with a small opening angle of $O(m'/m)$.  The direction 
of these particles point almost to the decay vertex, since the 
intermediate $s'$ is highly boosted with $\gamma \approx O(m/m')$.

The decay of $s'$ may be measured if it occurs inside a detector. 
To obtain a sufficient number of events, e.g.\ $N_{\rm min} \approx 
\mbox{a few}$, we need
\be
  n\, \varepsilon\, \sigma_{\rm SUSY} \Lint\,\, 
    {\rm min}\left\{ 1, \frac{L}{\gamma c\tau} \right\} 
    \simgt N_{\rm min},
\label{eq:detect}
\ee
where $n$, $\varepsilon$, $\sigma_{\rm SUSY}$, and $\Lint$ are the 
effective number of $s'$ per supersymmetric event, signal acceptance 
after cuts, total superparticle production cross section, and integrated 
luminosity, respectively.  The last factor in the left-hand-side 
represents a fraction of $s'$ decay occurring inside a detector with 
size $L$.  Taking $n \varepsilon \approx 0.1$, $\sigma_{\rm SUSY} \approx 
O(10~{\rm pb})$, and $\Lint \approx O(100~{\rm fb}^{-1})$, for example, 
\Eq{eq:detect} gives $\gamma c\tau \simlt O(10^{6}~\mbox{--}~10^7~{\rm cm}$) 
for $L \approx 1~{\rm m}$.  We therefore expect that the $s'$ decay 
can be seen at the LHC for a significant parameter region of $m' \simgt 
O(1~{\rm GeV})$, assuming that the background is under control.

If most of $s'$ decays inside a detector, i.e.\ $\gamma c\tau \simlt L$, 
then we may get a large number of $s'$ decay events $N_{\rm dec} = n\, 
\varepsilon\, \sigma_{\rm SUSY} \Lint$.  In this case, a rare 
decay mode into two photons may be observable if $N_{\rm dec} \simgt 
O(10^5~\mbox{--}~10^6)$.

For heavier $s'$ with $m' \simgt O(10~{\rm GeV})$, the $s'$ decays 
promptly.  For $m' \simlt 130~{\rm GeV}$, many supersymmetric 
events will be accompanied by one or two pairs of standard model 
particles---mostly $b\bar{b}$, $c\bar{c}$, or $\tau^+ \tau^-$, 
but also $\mu^+ \mu^-$ with $O(10^{-4})$ probability.  In contrast 
to the small $m'$ case, opening angles for these pairs are not very 
small.  For $m' \simgt 130~{\rm GeV}$, $s'$ decays (often dominantly) 
into a pair of electroweak gauge bosons (on- or off-shell), 
although it may also decay into $t\bar{t}$ or two Higgs bosons. 
The branching ratios into these modes depend on $\tan\beta$ and 
$\theta_{s' h_u}/\theta_{s' h_d}$.

Finally, we discuss decays of the CP odd component of $s'$.  The CP 
even and odd components of $s'$ typically have comparable decay lengths 
and branching ratios for $\Lambda_{\rm QCD} < m' \simlt 130~{\rm GeV}$ 
(except that the CP odd mixing angle $\theta_{s' h_u}$ has a $\cot\beta$ 
suppression at large $\tan\beta$).  For $m' \simgt 130~{\rm GeV}$, 
however, the CP odd component decays dominantly to either $b\bar{b}$, 
$t\bar{t}$, or the Higgs and $Z$ bosons, instead of two electroweak 
or Higgs bosons.  For $m' \simlt \Lambda_{\rm QCD}$, the leading 
hadronic decay of the CP odd component is to three rather than two 
mesons, and is thus suppressed by three-body phase space.

If CP is violated in the Higgs sector, the mass eigenstates are not the 
same as CP eigenstates.  In this case, both $s'$ mass eigenstates decay 
generically through the faster of the CP even and odd decay modes.

\section{Constraints}
\label{sec:constraints}

The hidden sector singlets and singlinos are additional neutral scalars 
and fermions which feebly interact with visible sector fields.  As such, 
they are constrained, albeit weakly, by existing experimental bounds 
from light axion, Higgs, and neutralino searches.

Our discussion will largely hinge on the magnitude of the mixing angles 
$\theta_{s' h_{u,d}}$ and $\theta_{\tilde{s}' \tilde h_{u,d}}$, as 
defined in \Eqs{eq:singlino-mix}{eq:singlet-mix}.  Parametrically, 
these mixing angles scale as
\bea
  && \theta_{s' h_{u,d}} \sim \theta_{\tilde{s}' \tilde h_{u,d}} 
    \sim \epsilon \frac{m'}{m} 
\nonumber\\
  && \quad \simeq 
    3 \times 10^{-5} \left( \frac{\epsilon}{10^{-2}} \right) 
    \left( \frac{m'}{1~{\rm GeV}} \right) 
    \left( \frac{300~{\rm GeV}}{m}\right).
\label{eq:mix-size}
\eea
Since they are naturally small, theories with singlet portal are 
constrained only weakly.

For $m'$ smaller than a few GeV, dominant constraints come from axion 
and light scalar searches.  In most cases, appropriate constraints 
can be estimated by replacing the axion decay constant $f_a$ in the 
axion bounds (for $\tan\beta \approx 1$) as
\be
  \frac{1}{f_a} \rightarrow {\rm max}\left\{ 
    \frac{\theta_{s' h_u}}{v_u}, \frac{\theta_{s' h_d}}{v_d} \right\}.
\label{eq:corresp}
\ee
For $m' \simlt 30~{\rm MeV}$, bounds from globular cluster 
stars, white dwarfs, and SN~1987A require $f_a \simgt 10^9~{\rm 
GeV}$~\cite{Raffelt:1990yz}.  We then find from \Eq{eq:corresp} 
that $\epsilon$ needs to be somewhat small, e.g.\ $\epsilon 
\simlt 10^{-3}$, for $m' \approx O(10~{\rm MeV})$.%
\footnote{Another possibility to evade the bounds for $m' \approx 
 O(10~{\rm MeV})$ is to have large $\epsilon \simgt 1/\tan\beta$, leading 
 to $f_a \simlt 10^6~{\rm GeV}$.  In this case, $s'$ produced in SN~1987A 
 is trapped inside the supernova, so that it does not carry away significant 
 energy.  Constraints from other astrophysical observations are irrelevant 
 for these values of $m'$.}
The complimentary regions in $m'$, either $\ll 10~{\rm MeV}$ or $\simgt 
100~{\rm MeV}$, is not constrained by astrophysics.

For $m' < m_K-m_\pi \simeq 350~{\rm MeV}$, rare processes such as $K^\pm 
\rightarrow \pi^\pm s'$ may give a constraint.  Since $s'$ in this 
mass range is long-lived, the corresponding axion bound is $f_a \simgt 
100~{\rm TeV}$~\cite{Mardon:2009gw}.  Given \Eqs{eq:mix-size}{eq:corresp}, 
however, this constraint is easily evaded.

For $2m_\mu < m' < m_B - m_K \simeq 4.8~{\rm GeV}$, the leading constraints 
come from rare decays of $B$ mesons, $B \rightarrow K s' \rightarrow 
K \mu^+ \mu^-$.  These decays place a stringent bound on the 
corresponding effective axion decay constant $f_a \simgt \mbox{few} 
\times 10^3~{\rm TeV}$~\cite{Batell:2009jf}.  Since the processes 
occur mainly with $s'$ emitted from internal top quarks, this translates 
into $\theta_{s' h_u} \simlt 10^{-4}$, obtained with $1/f_a$ replaced 
by $\theta_{s' h_u}/v_u$, rather than \Eq{eq:corresp}.  In view of 
\Eq{eq:mix-size}, this bound is satisfied in most of natural parameter 
regions.

If the hidden singlet is heavier, $m' > m_B - m_K$, then constraints 
may still arise from LEP results, e.g.\ from bounds on $s'$-strahlung 
and gauge boson fusion into $s'$.  However, the cross sections for 
these $s'$ production processes relative to the corresponding neutral 
Higgs boson production go roughly as $\theta_{s' h_{u,d}}^2$, so that 
they are typically very small.  Constraints from anomalous $Z$ boson 
decays are also easily satisfied, since $Z \rightarrow h^* s' \rightarrow 
b\bar{b}s'$ goes as $\theta_{s' h_{u,d}}^2$ while $Z \rightarrow s's'$ 
and $\tilde{s}'\tilde{s}'$ as $\theta_{s' h_{u,d}}^4$ and 
$\theta_{\tilde{s}' \tilde{h}_{u,d}}^4$, respectively.

\section{Dark Matter}
\label{sec:DM}

So far we have focused our attention on a simple hidden sector consisting 
of a single superfield $S'$.  In general, however, the hidden sector 
may have a much richer structure.  In particular, if it contains a 
particle which is stable on cosmological timescales, then that particle 
may comprise all of or a component of dark matter.  In this section, 
we discuss possible implications of such hidden sector dark matter.

As a simple example, let us consider the theory defined in \Eq{eq:simple-W}, 
augmented by $\mathbb{Z}_2$ stabilized dark matter, $H'$:
\be
  W_{\rm hid} = -\Lambda_{\rm eff}^2 S'+ \frac{\kappa'}{3} S'^3 
    + \frac{\lambda'}{2} S' H'^2.
\ee
This superpotential has a supersymmetry-preserving minimum at $\vev{S'} 
= \sqrt{\Lambda_{\rm eff}^2/\kappa'}$ and $\vev{H'} = 0$.  After 
supersymmetry breaking is communicated to the hidden sector, $\vev{S'}$ 
will shift but $\vev{H'}$ can still be vanishing.  The lightest 
component of $H'$ is then stable dark matter, whose mass, $m_{\rm DM}$, 
is given by the larger of $\lambda' \vev{S'} \approx O(\Lambda_{\rm eff})$ 
and gravity mediated contributions.

Which component of $\vev{H'}$ is the lightest depends on details of 
supersymmetry breaking.  In high-scale mediation scenarios, e.g.\ 
gravity mediation, both scalar and fermion components may receive 
sizable masses from supersymmetry breaking, which can be as large 
as the weak scale.  In gauge mediation, all components of $H'$ are 
nearly degenerate, as supersymmetry is approximately preserved in 
the hidden sector.  Small mass splittings, however, can be generated 
at order $\epsilon^2$.  The largest effect typically comes from 
$\vev{F_{S'}} \approx O(\epsilon^2 m^3/16\pi^2 m')$ induced by 
a supersymmetry breaking tadpole for $s'$, in which case dark matter 
is a real scalar field which is lighter than the other components by 
$O(\vev{F_{S'}}/m_{\rm DM})$.  While these mass splittings are small, 
typically of $O(100~{\rm keV}~\mbox{--}~10~{\rm GeV})$ for $\epsilon 
= 10^{-2}$, heavier components may still decay into lighter before 
today, depending on parameters (emitting either a gravitino or a pair 
of standard model particles).  If the decay into the gravitino is 
not kinematically allowed, then both the fermion and lighter scalar 
components are absolutely stable (due to $\mathbb{Z}_2 \times 
R\mbox{-parity}$).

The precise relic abundance of dark matter depends on the hidden 
sector spectrum, including mass splittings among components of 
$H'$.  For $\kappa' \simlt \lambda'$, it is roughly controlled by the 
thermally-averaged annihilation cross section into fields in the $S'$ 
multiplet $\vev{\sigma v} \approx \lambda'^4/64\pi m_{\rm DM}^2$, 
giving
\be
  \Omega_{\rm th} \approx 0.1 \left( \frac{0.1}{\lambda'} \right)^4 
    \left( \frac{m_{\rm DM}}{10~{\rm GeV}} \right)^2.
\label{eq:Omega}
\ee
Thus a stable particle(s) residing in the $H'$ multiplet can comprise 
all or a significant fraction of the dark matter in the universe.%
\footnote{We assume that $\tilde{s}'$ annihilates into $s'$ with a 
 sufficiently large cross section or decays into $s'$ and the gravitino 
 so that it does not overclose the universe.  Indeed, this condition 
 is satisfied in most of natural parameter regions.}

The scattering cross section of $H'$ dark matter with a target nucleus is 
dominated by $t$-channel exchange of CP even $s'$~\cite{Finkbeiner:2008qu}, 
and is given by
\be
  \sigma_T = \frac{\mu_T^2}{2\pi} \frac{\lambda'^2}{m'^4}
    \left( Z g_{s' pp} + (A-Z)g_{s' nn} \right)^2,
\label{eq:sigma_T}
\ee
for both fermion and scalar components, $h'$ and $\tilde{h}'$.  Here, 
$Z$ and $A$ are the atomic number and weight of the target nucleus, 
$\mu_T$ is the dark matter-nucleus reduced mass, and $g_{s' NN}$ for 
$N = p,n$ is the coupling of $s'$ to the proton and neutron:
\be
  g_{s' NN} = \frac{m_N}{\sqrt{2} v} 
    \left( \frac{\theta_{s' h_u}}{\sin\beta} \sum_{q=u,c,t} f^N_q 
    + \frac{\theta_{s' h_d}}{\cos\beta} \sum_{q=d,s,b} f^N_q \right).
\label{eq:g-sNN}
\ee
Using the following nucleon parameters~\cite{Drees:1993bu}
\be
\begin{array}{llll}
  f^p_u = 0.023, & f^p_d = 0.034, & f^p_s = 0.14, & f^p_{c,b,t} = 0.059, \\
  f^n_u = 0.019, & f^n_d = 0.041, & f^n_s = 0.14, & f^n_{c,b,t} = 0.059,
\end{array}
\label{eq:fNq}
\ee
we obtain the spin-independent dark-matter scattering cross section per 
nucleon, defined by $\sigma_T|_{A=Z=1}$,%
\footnote{An alternative definition for the dark matter-nucleon cross 
 section is sometimes used, see e.g.~\cite{Lewin:1995rx}, but the 
 difference is negligible for $g_{s' pp} \approx g_{s' nn}$ and the 
 dark matter sufficiently heavier than the nucleon.}
\bea
  \sigma &\simeq& 2 \times 10^{-48}~{\rm cm}^2\,\, 
    \frac{(1 + 1.6 \tan\beta)^2}{\sin^2\!\beta} 
    \left( \frac{\epsilon}{10^{-2}} \right)^2 
\nonumber\\
  && {} \times \left( \frac{\lambda'}{0.1} \right)^2 
    \left( \frac{10~{\rm GeV}}{m'} \right)^2 
    \left( \frac{300~{\rm GeV}}{m} \right)^2.
\label{eq:sigma}
\eea
Here, we have taken $\theta_{s' h_u} = \theta_{s' h_d} = \epsilon m'/m$ 
for illustrative purposes.  The cross section of \Eq{eq:sigma} is typically 
beyond the reach of current experiments, but it can be significantly 
enhanced at large $\tan\beta$ (and large $\lambda'$ and $\theta_{s' h_d}$).

\section{Discussion and Conclusions}
\label{sec:concl}

If the standard model is merely one of many sectors embedded in some 
fundamental theory---as is often the case in string theory---then these 
additional sectors may be  ``hidden'' in the sense that at low energies 
they interact only weakly with standard model particles.  In theories 
with weak scale supersymmetry, the characteristic mass scales associated 
with these hidden sectors can be naturally at or below the weak scale, 
generated through supersymmetry breaking effects.  Moreover, supersymmetry 
may offer a unique window into these hidden sectors via the decay of 
the LOSP, which can provide a rich phenomenology at colliders.

In this paper, we have considered a scenario in which the visible 
and hidden sectors both contain singlet chiral superfields, $S$ and $S'$, 
which, through a marginal kinetic mixing operator, connect the otherwise 
sequestered sectors.  This operator spontaneously induces a light mass 
scale $\Lambda_{\rm eff} \approx O(0.1~\mbox{--}~100~{\rm GeV})$ in 
the hidden sector.  Supersymmetric cascades necessarily produce Higgs 
bosons in an $O(0.01~\mbox{--}~1)$ fraction of events, and typically 
exhibit displaced decays of a light hidden sector state into standard 
model particles.

The theories discussed here may be easily discriminated from theories 
in which the lightest hidden sector particle is a hidden $U(1)$ gauge 
boson kinetically mixed with the photon.  If the mass of the hidden 
sector singlet is greater than the muon threshold, $m' > 2 m_\mu$, 
then the singlet portal gives the branching ratio to electrons versus 
muons ${\rm Br}(s' \rightarrow e^+ e^-)/{\rm Br}(s' \rightarrow 
\mu^+ \mu^-) \approx (m_e/m_\mu)^2 \ll 1$, while the hidden gauge boson 
case leads to a comparable branching ratio~\cite{ArkaniHamed:2008qp}. 
For $m' < 2 m_\mu$, the singlet portal yields a sizable decay rate 
to photons ${\rm Br}(s' \rightarrow \gamma\gamma) \simgt 10^{-3}$, 
while the rate to (three) photons is negligibly small in the gauge 
kinetic mixing scenario.

On the other hand, it may not be trivial to distinguish the singlet 
portal from a generic theory in which the lightest hidden sector 
particle is a scalar, $\phi'$.  In such a theory, the two-body decay 
of the lightest hidden sector particle typically yields heavy standard 
model fermions, ${\rm Br}(\phi' \rightarrow f\bar{f}) \propto m_f^2$, 
since this process requires a helicity flip in a final state fermion. 
However, in the case that the visible and hidden sectors are connected 
via a $U(1)$ gauge kinetic mixing---the only alternative to singlet 
kinetic mixing involving a marginal portal interaction---the two-body 
decay of $\phi'$ (via a loop of the hidden gauge boson, $\gamma'$, 
and $f$) is always accompanied by a four-body decay (via a tree diagram 
with two off-shell $\gamma'$).  The branching ratio of these two 
processes scales as ${\rm Br}(\phi' \rightarrow f\bar{f})/{\rm Br}(\phi' 
\rightarrow \gamma'^* \gamma'^* \rightarrow f\bar{f} f\bar{f}) \approx 
m_f^2 m_{\gamma'}^4 / m_{\phi'}^6$, so that the four-body decay rate 
may be significant, giving a different set of signatures for the 
portal out.  In more general cases, a detailed analysis of supersymmetric 
cascades may be required---for example, to discriminate from other 
theories, e.g.\ the ones considered in~\cite{Strassler:2006qa}.  In 
those cases, other characteristic features of the singlet portal, e.g.\ 
Higgs bosons arising from the end of the visible sector part of the 
cascades, will be important for identifying the underlying theory.

The analysis of the present paper can be straightforwardly extended to 
the case of multiple hidden sectors containing singlet chiral superfields, 
all kinetically mixed.  In this setup, visible sector superparticles 
typically cascade into the hidden sector which has the largest kinetic 
mixing with the visible sector.  The produced states may then cascade 
decay into states in another hidden sector or perhaps back to the 
visible sector.  These cascades will in general terminate at stable 
final states which are either hidden sector or standard model particles. 
Signatures of multiple hidden sector theories are, therefore, similar 
to the ones discussed in this paper.

Finally, let us present another class of singlet portal theories 
in which the hidden sector singlet is odd under $R$-parity.  We 
consider a hidden sector singlet $N'$ that kinetically mixes with 
right-handed neutrinos, $N$:
\be
  {\cal L} = \epsilon \int\!d^4\theta\, N^\dagger N' + \textrm{h.c.},
\label{eq:kinmix-N}
\ee
where we have omitted the generation index of $N$.  Here the fermionic 
components of $N \supset n$ and $N' \supset n'$ are the right-handed 
neutrino and the hidden sector singlino, respectively.  The right-handed 
neutrinos will have standard Majorana masses as well as neutrino 
Yukawa couplings
\be
  W = \frac{M}{2} N^2 + y_\nu L H_u N,
\label{eq:N}
\ee
leading to small neutrino masses, $m_\nu = y_\nu^2 v_u^2/M$, through 
the seesaw mechanism.  (Of course, Dirac neutrinos, $M = 0$ and 
$y_\nu \ll 1$, are also an option.)

In this scenario, the kinetic mixing terms in \Eq{eq:kinmix-N} do 
not induce an effective Polonyi term for the hidden sector, since the 
sleptons do not acquire VEVs.  Consequently, spontaneous generation 
of scales does not occur in theories of $R$-parity odd singlet 
kinetic mixing.

On the other hand, the neutrino portal in \Eq{eq:kinmix-N} can lead 
to distinctive signatures at colliders.  As in the case of $R$-parity 
even singlet kinetic mixing, supersymmetric cascades which originate 
in the visible sector invariably traverse into the hidden sector. 
This typically yields a Higgs, lepton, or neutrino at the bottom of 
the visible sector part of the cascades.  Meanwhile, any hidden sector 
singlino $n'$ produced in the process will decay back via $n' \rightarrow 
\ell^\pm h^\mp,\, \nu h$ with a macroscopic displacement, where $h^\pm$ 
and $h$ represent on or off-shell Higgs or electroweak gauge bosons. 
Whether these return processes indeed occur is model dependent, 
as in the case of other kinetic mixing portals.  For example, if 
$W_{\rm hid} \supset N' H'^2$ or $N'^3$, then the portal back may 
occur.  (The latter breaks $R$-parity.)  Here the mass scale of 
the hidden sector can be generated by gravity mediated contributions, 
which are of order or perhaps smaller than the weak scale.

The decay of $n'$ is mediated by operators induced via mass mixing 
between $n$ and $n'$, analogous to the $\tilde{s}$/$\tilde{s}'$ 
singlino case.  The mixing angles are
\be
  \theta_{n' n} \sim \epsilon \frac{m'}{M},
\label{eq:theta-n}
\ee
where $m'$ is the mass of $n'$.  This gives the decay width for 
$n' \rightarrow \ell^\pm W^\mp$
\be
  \Gamma_{n' \rightarrow \ell^\pm W^\mp} 
    \sim \frac{1}{8\pi} \frac{\epsilon^2 y_\nu^2 m'^3}{M^2} 
    = \frac{1}{8\pi} \frac{\epsilon^2 m_\nu m'^3}{v_u^2 M},
\label{eq:Gamma-n'}
\ee
assuming that the final state $W$ boson is on-shell.  (If not, the width 
is suppressed further.)  The decay length for the return process is then
\be
  \gamma c\tau \sim 10^7~{\rm cm} 
    \left( \frac{10^{-2}}{\epsilon} \right)^2 
    \left( \frac{200~{\rm GeV}}{m'} \right)^3 
    \left( \frac{M}{10^8~{\rm GeV}} \right),
\label{eq:ctau-n'}
\ee
where we have used $m_\nu = 0.1~{\rm eV}$.  This provides the 
possibility of observing the portal out process at the LHC for $M$ 
as large as $\approx O(10^8~{\rm GeV})$---close to the scale suggested 
by thermal leptogenesis~\cite{Buchmuller:2005eh}.  Unfortunately, the 
portal in process is often a slow, three-body decay, reducing the reach 
of $M$ by about two orders of magnitude, but the maximal reach can be 
obtained, e.g., if a sneutrino is the LOSP.  Note that the decay of $n'$ 
is not helicity suppressed, so the final state lepton can provide direct 
information on the flavor structure for $\epsilon$, $y_\nu$, and $M$.

More generally, visible return processes arising from supersymmetric 
hidden sector cascades can provide a unique and powerful probe of 
visible sector physics at very high energies.  The case of $R$-parity 
odd singlet kinetic mixing is a particular instance of employing highly 
displaced vertices to extend the reach of the LHC to extremely high 
energies.  Another example allowing for such a probe is given by 
a hidden sector $U(1)$ gauge field which kinetically mixes with a heavy 
$U(1)_{B-L}$ or $U(1)_{\rm flavor}$ gauge boson, but not with $U(1)$ 
hypercharge (although 
the reach is generically lower than the neutrino case).  In fact, 
this method can also apply to $R$-parity even singlet kinetic mixing, 
with the singlet $S$ having a large supersymmetric mass.

\begin{acknowledgments}
This work was supported in part by the Director, Office of Science, 
Office of High Energy and Nuclear Physics, of the US Department of 
Energy under Contract DE-AC02-05CH11231, and in part by the National 
Science Foundation under grants PHY-0555661 and PHY-0855653.
\end{acknowledgments}

\end{document}